\documentclass[12pt]{article}
\usepackage{amsmath}
\usepackage{color}
\usepackage{amssymb}
\usepackage{bm}

\makeatletter\@addtoreset{equation}{section}
\makeatother
\allowdisplaybreaks[2]
\begin{document}
\begin{titlepage}

\begin{flushright}
\phantom{preprint no.}
\end{flushright}
\vspace{0.5cm}
\begin{center}
{\Large \bf
Teukolsky-like equations with various spins\\
\vspace{2mm}
in a deformed Kerr spacetime
}
\lineskip .75em
\vskip0.5cm
{\large Hiroaki Nakajima${}^{1}$, Ya Guo${}^{2}$, and Wenbin Lin${}^{1,\,2,\,*}$}
\vskip 2.5em
${}^{1}$ {\normalsize\it School of Mathematics and Physics, \\
University of South China, Hengyang, 421001, China\\}
\vskip 1.0em
${}^{2}$ {\normalsize\it School of Physical Science and Technology, Southwest Jiaotong University, \\ Chengdu, 610031, China\\
}
\vskip 1.0em
${}^{*}$ {\normalsize\it Email: lwb@usc.edu.cn\\}
\vskip 1.0em
\vskip 3.0em
\end{center}
\begin{abstract}
We study the wave equations with the various spins on the background of the Kerr metric deformed by a function of the radial coordinate,
on which background we have studied the gravitational-wave equations previously.
We obtain the unified expression of the Teukolsky-like master equations and the corresponding radial equations with various spins.
We find that taking the separable gauge introduced in previous study simplifies the unified form.
We also discuss the structure of the radial equation as an ordinary differential equation, such as the existence of
the regular and the irregular singularities of the equation and the behavior of the solution around each singularity.

\end{abstract}
\end{titlepage}

\section{Introduction}
The study on the gravitational waves becomes much more important subject in cosmology than before, since the
direct observation of gravitational waves by LIGO and Virgo~\cite{Abbott:2016blz}.
The gravitational waves are described as a solution to the Einstein equation, where the metric is perturbed around a certain background.
As the source of the radiation of the gravitational waves, the dynamics of the binary system is often studied,
since the analytical calculation is possible as well as the numerical simulation.
In particular, if the system is for the case of the extreme mass-ratio inspiral (EMRI),
one can apply the  black-hole perturbation theory \cite{Mino:1997bx}, where the heavier object is regarded as the (black hole) background
and the lighter one is regarded as a test particle. if the mass ratio of the system is not so extreme, the background has to be deformed.
One systematic method to modify the background is proposed as
the effective one-body (EOB) dynamics \cite{Buonanno:2000ef,Damour:2001tu,Damour:2016gwp}.

The background of the EOB dynamics is deformed from the black hole spacetime, and then may not satisfy the vacuum Einstein equation.
However for some non-vacuum backgrounds which satisfy the Petrov type D condition \cite{Petrov}, the gravitational-wave equation can still be derived
using the Newman-Penrose formalism \cite{Newman:1961qr}, as for the Teukolsky equation \cite{Teukolsky:1973ha} in the vacuum case.
The advantage to use the Newman-Penrose formalism is that the role of the Einstein equation is rather restrictive, just used
to relate the Ricci tensor with the energy-momentum tensor. Then the extension to the non-vacuum case is relatively easier
\cite{Jing:2021ahx, Jing:2022vks, Guo:2023niy, Guo:2023hdn}.
For the spherically symmetric background, the type D condition is always satisfied.
A particular form of the background indeed appears in the EOB dynamics for the spinless binary system \cite{Buonanno:2000ef,Damour:2016gwp}.

On the other hand, for the axisymmetric background, the type D condition is not always satisfied.
In the previous paper \cite{Guo:2023wtx}, we have studied a certain axisymmetric background satisfying the type D condition
with some additional assumption, and have obtained the Kerr spacetime deformed by a function of the radial coordinate and the conformal factor.
Moreover, when this conformal factor is unity, we have derived the decoupled gravitational-wave equation.
Here the choice of the gauge condition has been important, since the separation of the variables is nontrivial \cite{Jing:2023vzq}.
We have found the \textit{separable gauge} such that the wave equation admits the separation of the variables.

In this paper we will study the massless wave equations with various spins (helicities, more precisely) as in \cite{Guo:2023hdn}
for the spherically symmetric background.
In order to avoid the complexities, we will first give the unified expression of those equations in terms of the Newman-Penrose quantities
as in \cite{Harris:2003eg, Vagenas:2020bys, Arbey:2021jif, Li:2011za}, where the the positive and the negative spins have been separately considered.

The remaining of this paper is organized as follows: in section 2, we will introduce our parametrization of the
background of the deformed Kerr spacetime, and show the quantities in the Newman-Penrose formalism.
In section 3, we will see the wave equation for spin $0$, $\pm 1/2$, $\pm 1$ and $\pm 2$ on this background.
Then we will give the unified expression for those equations with the general spin $s$.
We will also obtain the explicit Teukolsky-like master equation and the corresponding radial equation.
In section 4, we will discuss the structure of the radial equation, the existence of the regular and the irregular singularities,
and the behaviour of the solution around each singularity.
Section 5 is devoted to the summary and the discussion.

\section{Deformed Kerr metric and tetrads}

We will consider the gravitational background specified by the following metric \cite{Guo:2023wtx}:
\begin{align}
ds^{2}&=\frac{L(r)}{\Sigma}(dt-a\sin^{2}\theta d\varphi)^{2}-\frac{\Sigma}{L(r)}dr^{2}
\notag\\
&\qquad\qquad\quad {}
-\Sigma d\theta^{2}-\frac{\sin^{2}\theta}{\Sigma}\left[(r^{2}+a^{2})d\varphi-adt\right]^{2},
\label{metric1}
\end{align}
where $a$ is constant and $\Sigma$ is defined by
\begin{gather}
\Sigma=r^{2}+a^{2}\cos^{2}\theta.
\end{gather}
The function $L(r)$ is just restricted that it only depends on the radial coordinate $r$. If we assume the asymptotic flatness,
the large-$r$ behaviour is fixed as $L(r)\sim r^{2}$.
The metric \eqref{metric1} contains that of the Kerr spacetime as a special case as $L(r)=r^{2}-2Mr+a^{2}$.
The null tetrads corresponding to the metric can be taken as
\begin{align}
l=l_{\mu}dx^{\mu}&=dt-a\sin^{2}\theta d\varphi -\frac{\Sigma}{L}dr\ ,
\notag\\
n=n_{\mu}dx^{\mu}&=\frac{1}{2}\left[\frac{L}{\Sigma}(dt-a\sin^{2}\theta d\varphi) +dr\right],
\notag\\
m=m_{\mu}dx^{\mu}&=\frac{1}{\sqrt{2}\varrho}\left[-i(r^{2}+a^{2})\sin\theta d\varphi+ia\sin\theta dt-\Sigma d\theta\right],
\notag\\
\bar{m}=\bar{m}_{\mu}dx^{\mu}&=\frac{1}{\sqrt{2}\bar{\varrho}}\left[i(r^{2}+a^{2})\sin\theta d\varphi- ia\sin\theta dt-\Sigma d\theta\right],
\label{tetrads1}
\end{align}
where $\varrho=r+ia\cos\theta$ and the null tetrads satisfy
\begin{gather}
ds^{2}=2ln-2m\bar{m}.
\label{decomp}
\end{gather}
From \eqref{tetrads1}, one can compute the spin coefficients, the components
of the Ricci tensor and the Weyl scalars as \cite{Guo:2023wtx}
\begin{gather}
\kappa=\lambda=\sigma=\nu=0,
\label{GS}
\\
\epsilon=0,
\quad
\rho=-\frac{1}{\bar{\varrho}},
\quad
\tau=-\frac{ia\sin\theta}{\sqrt{2}\Sigma},
\quad
\mu=-\frac{L}{2\bar{\varrho}\Sigma},
\\
\gamma=\mu+\frac{L'}{4\Sigma},
\quad
\pi=\frac{ia\sin\theta}{\sqrt{2}\bar{\varrho}^{2}},
\quad
\beta=\frac{\cot\theta}{2\sqrt{2}\varrho},
\quad
\alpha=\pi-\bar{\beta},
\\
\Psi_{0}=\Psi_{1}=\Psi_{3}=\Psi_{4}=0,
\label{typeD}
\\
\Psi_{2}=\frac{1}{\Sigma}\left(
-\frac{1}{6}-\frac{a^{2}\sin^{2}\theta}{\bar{\varrho}^{2}}-\frac{ia\cos\theta}{\bar{\varrho}}+\frac{L}{\bar{\varrho}^{2}}
-\frac{L'}{2\bar{\varrho}}+\frac{L''}{12}
\right),
\\
\Phi_{00}=\Phi_{01}=\Phi_{10}=\Phi_{02}=\Phi_{20}=\Phi_{12}=\Phi_{21}=0,
\label{simple}
\\
\Phi_{11}=\frac{1}{4\Sigma}
\left(\frac{2r^{2}-2a^{2}+2L-2rL'}{\Sigma}+\frac{L''}{2}-1\right),
\label{simple1}
\\
\Lambda=\frac{2-L''}{24\Sigma}.
\label{simple2}
\end{gather}
where we follow the notations of Newman-Penrose \cite{Newman:1961qr} and Pirani \cite{Pirani}.
The same notation is also used in Teukolsky \cite{Teukolsky:1973ha}.
The prime denotes the derivative with respect to $r$.
We note that the equation \eqref{typeD} implies that the background belongs the Petrov type D,
which will be important to derive the wave equation on this background.
We also have \eqref{GS}, which follows from the non-vacuum extension \cite{KT, RS} of the Goldberg-Sachs theorem \cite{GS}
(see also \cite{Stephani:2003tm}).
We also note that the conformal transformation of the metric
\begin{gather}
d\tilde{s}^{2}=A(r,\theta)ds^{2}
\label{metric2}
\end{gather}
also satisfies \eqref{GS} and the type D condition \eqref{typeD} \cite{Guo:2023wtx}. However \eqref{simple} is no longer satisfied in general,
which makes complicated to obtain the gravitational-wave equation. Then we here consider the case $A(r,\theta)=1$ only.
As an exceptional case, for $A(r,\theta)=A(r)$ and $a=0$, the metric \eqref{metric2} reduces to the general form
of the spherically symmetric metric, and we have obtained the wave equations for general $A(r)$ \cite{Guo:2023hdn}
with the different radial coordinate.
In this case, just $\Phi_{11}$, $\Lambda$, $\Phi_{00}$ and $\Phi_{22}$ are nonvanishing in general.

\section{Wave equations with various spins}

We will consider the wave equation with the various spin $s$ on the background in the previous section.
For simplicity, we will assume that the back reactions from the matters and the electromagnetic fields to the gravitational background are negligible.
and all the fields are minimally coupled to the background of the gravitational field, unless it is specified.

\subsection{spin $0$}

The massless Klein-Gordon equation in the gravitational background is
\begin{gather}
\Box\phi=\nabla_{\mu}(g^{\mu\nu}\partial_{\nu}\phi)=\frac{1}{\sqrt{-g}}\partial_{\mu}(\sqrt{-g}g^{\mu\nu}\partial_{\nu}\phi)=T,
\label{KG1}
\end{gather}
where $T$ is the source.
$\nabla_{\mu}$ is the covariant derivative with respect to the curved spacetime (not for the local Lorentz transformation),
of which the projection by the null tetrads gives
\begin{gather}
D=l^{\mu}\nabla_{\mu}, \quad \Delta=n^{\mu}\nabla_{\mu}, \quad \delta=m^{\mu}\nabla_{\mu}, \quad
\bar{\delta}=\bar{m}^{\mu}\nabla_{\mu}.
\label{diffop}
\end{gather}
By decomposing $g^{\mu\nu}$ in terms of the null tetrads, \eqref{KG1} can be rewritten as
\begin{align}
&\bigl[(\Delta-\gamma-\bar{\gamma}+\mu+\bar{\mu})D+(D+\epsilon+\bar{\epsilon}-\rho-\bar{\rho})\Delta
\notag\\
&\qquad {}-(\bar{\delta}-\alpha+\bar{\beta}+\pi-\bar{\tau})\delta-(\delta-\bar{\alpha}+\beta+\bar{\pi}-\tau)\bar{\delta}\bigr]\phi=T,
\label{KG2}
\end{align}
where
we have used the relation for the spin coefficients as
\begin{align}
\nabla_{\mu}l^{\mu}&=\epsilon+\bar{\epsilon}-\rho-\bar{\rho}, &
\nabla_{\mu}n^{\mu}&=-\gamma-\bar{\gamma}+\mu+\bar{\mu},
\notag\\
\nabla_{\mu}m^{\mu}&=-\bar{\alpha}+\beta+\bar{\pi}-\tau, &
\nabla_{\mu}\bar{m}^{\mu}&=-\alpha+\bar{\beta}+\pi-\bar{\tau}.
\end{align}
For later convenience, we will rewrite \eqref{KG2} such that the order of the differential operators is rearranged,
which satisfy the commutation relations
\begin{align}
\Delta D - D \Delta&=
(\gamma+\bar{\gamma}) D +(\epsilon+\bar{\epsilon})\Delta-(\pi+\bar{\tau})\delta-(\bar{\pi}+\tau)\bar{\delta},
\notag\\
\delta\bar{\delta}-\bar{\delta}\delta&=
(\mu-\bar{\mu})D+(\rho-\bar{\rho})\Delta+(\bar{\alpha}-\beta)\bar{\delta}+(\alpha-\bar{\beta})\delta.
\label{com0}
\end{align}
We will also use
\begin{gather}
\Delta\rho-\bar{\delta}\tau=
\nu\kappa-\sigma\lambda+(\gamma+\bar{\gamma}-\bar{\mu})\rho-(\alpha-\bar{\beta}+\bar{\tau})\tau-\Psi_{2}-2\Lambda,
\label{NP1}
\end{gather}
which is one of the Newman-Penrose equation.
Then \eqref{KG2} can be rewritten as
\begin{gather}
\left[(\Delta-\gamma-\bar{\gamma}+\bar{\mu})(D-\rho)-(\bar{\delta}-\alpha+\bar{\beta}-\bar{\tau})(\delta-\tau)-\Psi_{2}-2\Lambda\right]\phi=
\frac{T}{2},
\label{KG3}
\end{gather}
We note that the last term $-2\Lambda\phi$ in the right hand side is responsible to the minimal coupling.
If we consider the curvature coupling, there is the contribution $c\mathcal{R}\phi=24c\Lambda\phi$,
where $\mathcal{R}$ is the Ricci scalar and $c$ is constant.

\subsection{spin $\pm 1/2$}

The Weyl equation in the gravitational background in the Newman-Penrose formalism is written as
\begin{align}
(\bar{\delta}-\alpha+\pi)\chi_{0}-(D-\rho+\epsilon)\chi_{1}=0,
\label{weyl1}
\\
(\Delta-\gamma+\mu)\chi_{0}-(\delta+\beta-\tau)\chi_{1}=0,
\label{weyl2}
\end{align}
where $\chi_{0}$ and $\chi_{1}$ are the components of the Weyl spinor.
One can eliminate $\chi_{0}$ using the following commutation relation:
\begin{align}
&\left[\Delta+(p+1)\gamma-\bar{\gamma}-q\mu+\bar{\mu}\right](\bar{\delta}+p\alpha-q\pi) \notag\\
&\qquad -\left[\bar{\delta}+(p+1)\alpha+\bar{\beta}-\bar{\tau}-q\pi\right](\Delta+p\gamma-q\mu) \notag\\
&=\nu D -\lambda\delta-p\left[(\beta+\tau)\lambda-(\rho+\epsilon)\nu+\Psi_{3}\right] \notag\\
&\qquad +q\left[-D\nu+\delta\lambda+(\bar{\pi}+\tau+3\beta-\bar{\alpha})\lambda
-(3\epsilon+\bar{\epsilon}+\rho-\bar{\rho})\nu+2\Psi_{3}\right] \notag\\
&=0, \label{com1}
\end{align}
where $p$ and $q$ are arbitrary constants and we have just used $\nu=\lambda=\Psi_{3}=0$ for the last equality.
Hence \eqref{com1} holds not only on the vacuum, but also on the current background.
We will obtain the wave equation for $\chi_{1}$ in a similar way with the method
used to derive the Teukolsky equation \cite{Teukolsky:1973ha}.
We operate $(\Delta-\gamma+\mu+\bar{\mu})$ on \eqref{weyl1} and $(\bar{\delta}+\bar{\beta}+\pi-\bar{\tau})$ on \eqref{weyl2}, respectively.
We then compute the difference of them. The terms with $\chi_{0}$ are canceled from \eqref{com1} with $p=q=-1$,
and the remaining part is
\begin{gather}
\bigl[(\Delta-\bar{\gamma}+\mu+\bar{\mu})(D+\epsilon-\rho)-(\bar{\delta}+\bar{\beta}+\pi-\bar{\tau})(\delta+\beta-\tau)\bigr]\chi_{1}=0,
\label{chi1}
\end{gather}
which gives the wave equation for $s=-1/2$. In a similar way, the wave equation of $\chi_{0}$ (for $s=1/2$) is obtained as
\begin{gather}
\bigl[(D+\bar{\epsilon}-\rho-\bar{\rho})(\Delta-\gamma+\mu)-(\delta-\bar{\alpha}-\tau+\bar{\pi})(\bar{\delta}-\alpha+\pi)\bigr]\chi_{0}=0.
\label{chi0-1}
\end{gather}
We note that \eqref{chi1} and \eqref{chi0-1} take the same form as the vacuum case \cite{Teukolsky:1973ha}.
For later convenience, we rewrite \eqref{chi0-1} as in the previous subsection. We will use \eqref{com0}, \eqref{NP1} and
\begin{align}
D\gamma-\Delta\epsilon&=\alpha(\tau+\bar{\pi})+\beta(\bar{\tau}+\pi)-\gamma(\epsilon+\bar{\epsilon})-\epsilon(\gamma+\bar{\gamma})+\tau\pi-\nu\kappa+\!\Psi_{2}\!+\!\Phi_{11}\!-\!\Lambda,
\label{NP2}
\\
\delta\alpha-\bar{\delta}\beta&=\mu\rho-\lambda\sigma+\alpha\bar{\alpha}+\beta\bar{\beta}-2\alpha\beta+\gamma(\rho-\bar{\rho})+\epsilon(\mu-\bar{\mu})
-\Psi_{2}+\Phi_{11}+\Lambda,
\label{NP3}
\\
D\mu-\delta\pi&=\pi(\bar{\pi}-\bar{\alpha}+\beta)-\mu(\epsilon+\bar{\epsilon}-\bar{\rho})-\nu\kappa+\lambda\sigma+\Psi_{2}+2\Lambda.
\label{NP4}
\end{align}
Then \eqref{chi0-1} can be rewritten as
\begin{gather}
\bigl[(\Delta-2\gamma-\bar{\gamma}+\bar{\mu})(D-\epsilon-2\rho)
-(\bar{\delta}-2\alpha+\bar{\beta}-\bar{\tau})(\delta-\beta-2\tau)-3\Psi_{2}\bigr]\chi_{0}=0,
\label{chi0-2}
\end{gather}

\subsection{spin $\pm 1$}

The Maxwell equation in the gravitational background is
\begin{align}
(D-2\rho)\phi_{1}-(\bar{\delta}+\pi-2\alpha)\phi_{0}&=J_{l},
\label{max1}
\\
(\delta-2\tau)\phi_{1}-(\Delta+\mu-2\gamma)\phi_{0}&=J_{m},
\label{max2}
\\
(D-\rho+2\epsilon)\phi_{2}-(\bar{\delta}+2\pi)\phi_{1}&=J_{\bar{m}},
\label{max3}
\\
(\delta-\tau+2\beta)\phi_{2}-(\Delta+2\mu)\phi_{1}&=J_{n},
\label{max4}
\end{align}
where $\phi_{0}$, $\phi_{1}$ and $\phi_{2}$ are complex, and are constructed from the field strength (the Faraday tensor) $F_{\mu\nu}$ as
\begin{gather}
\phi_{0}=F_{\mu\nu}l^{\mu}m^{\nu}, \quad
\phi_{1}=\frac{1}{2}F_{\mu\nu}(l^{\mu}n^{\nu}+\bar{m}^{\mu}m^{\nu}), \quad
\phi_{2}=F_{\mu\nu}\bar{m}^{\mu}n^{\nu}.
\end{gather}
$J_{l}$, $J_{n}$, $J_{m}$ and $J_{\bar{m}}$ are the projection of the current $J^{\mu}$ by the null tetrads as $J_{l}=J^{\mu}l_{\mu}$, etc.
As in the vacuum case, we will consider the wave equation just for $\phi_{0}$ and $\phi_{2}$, from which $\phi_{1}$ can be obtained
using \eqref{max1}--\eqref{max4}.
One can make the wave equation for $\phi_{2}$ by eliminating $\phi_{1}$ from \eqref{max3} and \eqref{max4}
as a similar procedure used in the previous subsection.
Using the commutation relation \eqref{com1} with $p=0$ and $q=-2$, we have
\begin{gather}
\bigl[(\Delta+\gamma-\bar{\gamma}+2\mu+\bar{\mu})(D+2\epsilon-\rho)-(\bar{\delta}+\alpha+\bar{\beta}+2\pi-\bar{\tau})(\delta+2\beta-\tau)\bigr]
\phi_{2}=J_{2},
\label{phi2}
\end{gather}
where $J_{2}$ is defined by
\begin{gather}
J_{2}=(\Delta+\gamma-\bar{\gamma}+2\mu+\bar{\mu})J_{\bar{m}}-(\bar{\delta}+\alpha+\bar{\beta}+2\pi-\bar{\tau})J_{n}.
\end{gather}
In a similar way, one can obtain the wave equation for $\phi_{0}$ from \eqref{max1} and \eqref{max2} as
\begin{gather}
\bigl[(D-\epsilon+\bar{\epsilon}-2\rho-\bar{\rho})(\Delta-2\gamma+\mu)-(\delta-\beta-\bar{\alpha}-2\tau+\bar{\pi})(\bar{\delta}+\pi-2\alpha)\bigr]
\phi_{0}=J_{0},
\label{phi0-1}
\end{gather}
where $J_{0}$ is defined by
\begin{gather}
J_{0}=(\delta-\beta-\bar{\alpha}-2\tau+\bar{\pi})J_{l}-(D-\epsilon+\bar{\epsilon}-2\rho-\bar{\rho})J_{m}.
\end{gather}
We note that \eqref{phi2} and \eqref{phi0-1} take the same form as the vacuum case \cite{Teukolsky:1973ha}.
For later convenience, we rewrite \eqref{phi0-1} using \eqref{com0}, \eqref{NP1} and \eqref{NP2}--\eqref{NP4} as
\begin{align}
&\bigl[(\Delta-3\gamma-\bar{\gamma}+\bar{\mu})(D-2\epsilon-3\rho)
\notag\\
&\qquad {}-(\bar{\delta}-3\alpha+\bar{\beta}-\bar{\tau})(\delta-2\beta-3\tau)-6\Psi_{2}\bigr]\phi_{0}=J_{0},
\label{phi0-2}
\end{align}
We note that as in the case of $s=\pm 1/2$, the equations \eqref{phi2} and \eqref{phi0-2} take the same form
as those on the vacuum background \cite{Teukolsky:1973ha}.

\subsection{spin $\pm 2$}

The wave equation for the spin $\pm 2$ is obtained from the perturbed Einstein equation or the perturbed Newman-Penrose equation.
We will here use the latter formalism, because it is easier than the former to consider the non-vacuum background.
In the previous studies, it has been found that the gravitational-wave equation on the non-vacuum background depends on the gauge.
We take the \textit{separable gauge} \cite{Guo:2023wtx} as
\begin{align}
&\frac{1}{2}(\Phi_{11}-3\Lambda)(\Delta+3\gamma-\bar{\gamma}+\mu+\bar{\mu})\lambda^{B}
+[(\Delta+\mu+\bar{\mu})\Phi_{11}]\lambda^{B}
\notag\\
&\quad {}+\frac{1}{2}(\Phi_{11}+3\Lambda)(\bar{\delta}+3\alpha+\bar{\beta}-\bar{\tau}+\pi)\nu^{B}
+[(\bar{\delta}+\pi-\bar{\tau})\Phi_{11}]\nu^{B}=0,
\label{gauge03}
\end{align}
where the superscript $B$ denotes the perturbation part of the corresponding quantities. The superscript $A$ for the background part has been
omitted for the notational simplicity.
%
Under the separable gauge, the wave equation for the perturbation part of the Weyl scalar $\Psi_{4}^{B}$ is
\begin{align}
&\bigl[(\Delta+3\gamma-\bar{\gamma}+4\mu+\bar{\mu})(D+4\epsilon-\rho)
\notag\\
&\qquad {}-(\bar{\delta}+3\alpha+\bar{\beta}-\bar{\tau}+4\pi)(\delta+4\beta-\tau)-3\Psi_{2}-6\Lambda\bigr]\Psi_{4}^{B}=T_{4},
\label{GW04}
\end{align}
where the source $T_{4}$ is defined by
\begin{align}
T_{4}&=(\Delta+3\gamma-\bar{\gamma}+4\mu+\bar{\mu})
\bigl[(\bar{\delta}-2\bar{\tau}+2\alpha)\Phi_{21}^{B}-(\Delta+2\gamma-2\bar{\gamma}+\bar{\mu})\Phi_{20}^{B}\bigr]
\notag\\
&\qquad {}-(\bar{\delta}+3\alpha+\bar{\beta}-\bar{\tau}+4\pi)
\bigl[(\bar{\delta}+2\alpha+2\bar{\beta}-\bar{\tau})\Phi_{22}^{B}-(\Delta+2\gamma+2\bar{\mu})\Phi_{21}^{B})\bigr].
\end{align}
In a similar way, one can also obtain the wave equation for $\Psi_{0}^{B}$ as
\begin{align}
&\bigl[(D-3\epsilon+\bar{\epsilon}-4\rho-\bar{\rho})(\Delta-4\gamma+\mu)
\notag\\
&\qquad {}-(\delta-\bar{\alpha}-3\beta-4\tau+\bar{\pi})(\bar{\delta}+\pi-4\alpha)-3\Psi_{2}-6\Lambda\bigr]\Psi_{0}^{B}=T_{0},
\label{GW00-1}
\end{align}
where the source $T_{0}$ is defined by
\begin{align}
T_{0}&=(\delta-\bar{\alpha}-3\beta-4\tau+\bar{\pi})
\bigl[(D-2\epsilon-2\rho)\Phi_{01}^{B}-(\delta-2\bar{\alpha}-2\beta+\bar{\pi})\Phi_{00}^{B}\bigr]
\notag\\
&\qquad {}-(D-3\epsilon+\bar{\epsilon}-4\rho-\bar{\rho})
\bigl[(D-2\epsilon+2\bar{\epsilon}-\bar{\rho})\Phi_{02}^{B}-(\delta-2\beta+2\bar{\pi})\Phi_{01}^{B}\bigr].
\end{align}
The condition for the separable gauge is
\begin{align}
&\frac{1}{2}(\Phi_{11}-3\Lambda)(D-3\epsilon+\bar{\epsilon}-\rho-\bar{\rho})\sigma^{B}
+[(D-\rho-\bar{\rho})\Phi_{11}]\sigma^{B}
\notag\\
&\quad {}+\frac{1}{2}(\Phi_{11}+3\Lambda)(\delta+\bar{\pi}-\bar{\alpha}-3\beta-\tau)\kappa^{B}
+[(\delta+\bar{\pi}-\tau)\Phi_{11}]\kappa^{B}=0,
\label{gauge04}
\end{align}
For later convenience, we rewrite \eqref{GW00-1} using \eqref{com0}, \eqref{NP1} and \eqref{NP2}--\eqref{NP4} as
\begin{align}
&\bigl[(\Delta-5\gamma-\bar{\gamma}+\bar{\mu})(D-4\epsilon-5\rho)
\notag\\
&\qquad {}-(\bar{\delta}-5\alpha+\bar{\beta}-\bar{\tau})(\delta-4\beta-5\tau)-15\Psi_{2}-6\Lambda\bigr]\Psi_{0}^{B}=T_{0}.
\label{GW00-2}
\end{align}

\subsection{unified wave equation for general spin $s$}

We can unify the above wave equations \eqref{KG3}, \eqref{chi1}, \eqref{chi0-2}, \eqref{phi2}, \eqref{phi0-2}, \eqref{GW04} and \eqref{GW00-2},
and can express the unified one for the general spin $s$ as
\begin{align}
&\biggl\{\bigl[\Delta-(2s+1)\gamma-\bar{\gamma}+(|s|-s)\mu+\bar{\mu}\bigr]\bigl[D-2s\epsilon-(|s|+s+1)\rho\bigr]
\notag\\
&\qquad {}-\bigl[\bar{\delta}-(2s+1)\alpha+\bar{\beta}-\bar{\tau}+(|s|-s)\pi\bigr]\bigl[\delta-2s\beta-(|s|+s+1)\tau\bigr]
\notag\\
&\qquad {}-(1+3s+2s^{2})\Psi_{2}-2(1-3|s|+2|s|^{2})\Lambda\biggr\}\tilde{\psi}_{(s)}=\tilde{T}_{(s)},
\label{unified1}
\end{align}
where we have collectively denoted the fields and the sources as
\begin{gather}
\tilde{\psi}_{(s)}=
\begin{cases}
\Psi_{4}^{B} & \text{for $s=-2$}
\\
\phi_{2} & \text{for $s=-1$}
\\
\chi_{1} & \text{for $s=-1/2$}
\\
\phi & \text{for $s=0$}
\\
\chi_{0} & \text{for $s=1/2$}
\\
\phi_{0} & \text{for $s=1$}
\\
\Psi_{0}^{B} & \text{for $s=2$}
\end{cases},
\qquad
\tilde{T}_{(s)}=
\begin{cases}
T_{4} & \text{for $s=-2$}
\\
J_{2} & \text{for $s=-1$}
\\
0 & \text{for $s=-1/2$}
\\
T/2 & \text{for $s=0$}
\\
0 & \text{for $s=1/2$}
\\
J_{0} & \text{for $s=1$}
\\
T_{0} & \text{for $s=2$}
\end{cases}.
\end{gather}
We will rewrite \eqref{unified1} to the slightly simpler form by the following redefinition:
\begin{gather}
\psi_{(s)}=\exp\bigl[(|s|-s)f\bigr]\tilde{\psi}_{(s)}, \quad T_{(s)}=\exp\bigl[(|s|-s)f\bigr]\tilde{T}_{(s)},
\label{redef}
\end{gather}
where $f$ is a function of $r$ and $\theta$, and will be determined soon. By substituting \eqref{redef} into \eqref{unified1}, we have
\begin{align}
&\biggl\{\bigl[\Delta-(2s+1)\gamma-\bar{\gamma}+\bar{\mu}+(|s|-s)(\mu-\Delta f)\bigr]\bigl[D-2s\epsilon-(2s+1)\rho-(|s|-s)(\rho+Df)\bigr]
\notag\\
&\quad {}-\bigl[\bar{\delta}-(2s+1)\alpha+\bar{\beta}-\bar{\tau}+(|s|-s)(\pi-\bar{\delta}f)\bigr]
\bigl[\delta-2s\beta-(2s+1)\tau-(|s|-s)(\tau+\delta f)\bigr]
\notag\\
&\quad {}-(1+3s+2s^{2})\Psi_{2}-2(1-3|s|+2|s|^{2})\Lambda\biggr\}\psi_{(s)}=T_{(s)},
\label{unified1-1}
\end{align}
One can find that $\mu-\Delta f$,  $\rho+Df$, $\pi-\bar{\delta}f$ and $\tau+\delta f$ can simultaneously vanish
if the following two equations are satisfied:
\begin{gather}
\partial_{r}f=\frac{1}{\bar{\varrho}},\quad \partial_{\theta}f=\frac{ia\sin\theta}{\bar{\varrho}}.
\end{gather}
The above can be integrated as $f=\ln\bar{\varrho}$, 
and hence
\begin{gather}
\psi_{(s)}= \bar{\varrho}^{|s|-s}\tilde{\psi}_{(s)},\quad T_{(s)}= \bar{\varrho}^{|s|-s}\tilde{T}_{(s)}.
\label{redef1}
\end{gather}
Then \eqref{unified1-1} is simplified as
\begin{align}
&\biggl\{\bigl[\Delta-(2s+1)\gamma-\bar{\gamma}+\bar{\mu}\bigr]\bigl[D-2s\epsilon-(2s+1)\rho\bigr]
\notag\\
&\qquad {}-\bigl[\bar{\delta}-(2s+1)\alpha+\bar{\beta}-\bar{\tau}\bigr]\bigl[\delta-2s\beta-(2s+1)\tau\bigr]
\notag\\
&\qquad {}-(1+3s+2s^{2})\Psi_{2}-2(1-3|s|+2|s|^{2})\Lambda\biggr\}\psi_{(s)}=T_{(s)}.
\label{unified2}
\end{align}
The advantage of the above form is that
after the transformation \eqref{redef1}, $|s|$-dependence appears only in the coefficient of $\Lambda$ which vanishes on the vacuum.
We note that the same transformation has been performed in the vacuum case as well \cite{Teukolsky:1973ha}.

By substituting the background, the explicit form of the unified wave equation \eqref{unified2} is
\begin{align}
&\left[\frac{(r^{2}+a^{2})^{2}}{L(r)}-a^{2}\sin^{2}\theta\right]\frac{\partial^{2}\psi_{(s)}}{\partial t^{2}}
-s\left[\frac{(r^{2}+a^{2})L'(r)}{L(r)}-4r-2ia\cos\theta\right]\frac{\partial\psi_{(s)}}{\partial t}
\notag\\
&{}-L^{-s}(r)\frac{\partial}{\partial r}\left(L^{s+1}(r)\frac{\partial\psi_{(s)}}{\partial r}\right)
-\frac{1}{\sin\theta}\frac{\partial}{\partial\theta}\left(\sin\theta\frac{\partial\psi_{(s)}}{\partial\theta}\right)
+\left(\frac{a^{2}}{L(r)}-\frac{1}{\sin^{2}\theta}\right)\frac{\partial^{2}\psi_{(s)}}{\partial\varphi^{2}}
\notag\\
&{}+\left[\frac{2a(r^{2}+a^{2})}{L(r)}-2a\right]\frac{\partial^{2}\psi_{(s)}}{\partial t \partial\varphi}
-s\left(\frac{aL'(r)}{L(r)}+\frac{2i\cos\theta}{\sin^2\theta}\right)\frac{\partial\psi_{(s)}}{\partial\varphi}
\notag\\
&{}+\left[s^{2}\cot^{2}\theta-s-\frac{1}{2}(|s|+s)(L''(r)-2)\right]\psi_{(s)}=2\Sigma T_{(s)},
\label{PDE1}
\end{align}
We note that \eqref{PDE1} with $s=\pm 2$ looks different from the equation obtained in our previous study \cite{Guo:2023niy}.
However this is just due to the difference of the transformation \eqref{redef1}, and they are equivalent.
We also note that in the case of $L(r)=r^{2}-2Mr+a^{2}$, \eqref{PDE1} reduces to the Teukolsky master equation
with the spin $s$ on the background of the Kerr spacetime.
For the homogeneous case. the equation \eqref{PDE1} allows the separation of the variables, and we assume the product form of the solution as
\begin{gather}
\psi_{(s)}=e^{-i\omega t}e^{im\varphi}R(r)S(\theta),
\end{gather}
where $\omega$ is the frequency of the waves and $m$ is constant.
Then the separated equations are
\begin{align}
&L^{-s}\frac{d}{dr}\left(L^{s+1}\frac{dR}{dr}\right)
\notag\\
&\qquad {}
+\biggl[\frac{K^{2}-isKL'}{L}+\frac{1}{2}(|s|+s)(L''-2)+4is\omega r+2ma\omega-a^{2}\omega^{2}-\boldsymbol{\lambda}_{(s)}\biggr]R=0,
\label{radial}
\\
&\frac{1}{\sin\theta}\frac{d}{d\theta}\left(\sin\theta\frac{dS}{d\theta}\right)
\notag\\
&{}\qquad +\left(a^{2}\omega^{2}\cos^{2}\theta-\frac{m^{2}}{\sin^{2}\theta}-2sa\omega\cos\theta-\frac{2ms\cos\theta}{\sin^{2}\theta}
-s^{2}\cot^2\theta+s+\boldsymbol{\lambda}_{(s)}\right)S=0,
\label{angular}
\end{align}
where $K=(r^{2}+a^{2})\omega-ma$ and $\boldsymbol{\lambda}_{(s)}$ is the separation constant.
For the positive $s$,  \eqref{radial} has been derived for the spherically symmetric background \cite{Harris:2003eg},
for which our result gives the natural extension.
As argued also in \cite{Harris:2003eg}, $L''$-term in \eqref{radial} can be removed by the transformation $R=L^{-s}\hat{R}$ for positive $s$,
which can be generalized into general $s$ as
\begin{gather}
R=L^{-\frac{1}{2}(|s|+s)}\hat{R},
\label{redef2}
\end{gather}
and then $\hat{R}$ satisfies
\begin{gather}
L^{|s|}\frac{d}{dr}\left(L^{-|s|+1}\frac{d\hat{R}}{dr}\right)
+\biggl(\frac{K^{2}-isKL'}{L}+4is\omega r-\hat{\boldsymbol{\lambda}}_{(s)}\biggr)\hat{R}=0,
\label{radial0}
\end{gather}
where $\hat{\boldsymbol{\lambda}}_{(s)}$ is defined by
\begin{gather}
\hat{\boldsymbol{\lambda}}_{(s)}=\boldsymbol{\lambda}_{(s)}+s+|s|-2ma\omega+a^{2}\omega^{2}.
\end{gather}
From \eqref{angular}, one can find that $S(\theta)e^{im\varphi}$ coincides with the spin-weighted spheroidal harmonics
${}^{}_{s}S^{a\omega}_{lm}(\theta,\varphi)$ with the spin $s$,
where $l$ and $m$ take the values of
\begin{gather}
l=|s|,\ |s|+1,\ |s|+2,\ \ldots, \quad m= -l,\ -l+1,\ \ldots,\ l-1,\ l,
\end{gather}
respectively.
$\boldsymbol{\lambda}_{(s)}$ becomes the eigenvalue of ${}^{}_{s}S^{a\omega}_{lm}(\theta,\varphi)$, of which small-$a\omega$ expansion has been
computed as \cite{Seidel:1988ue, Berti:2005gp}
\begin{gather}
\boldsymbol{\lambda}_{(s)}=(l-s)(l+s+1)-\frac{2s^{2}m}{l(l+1)}a\omega+\left[h(l+1)-h(l)-1\right]a^{2}\omega^{2}+O(a^{3}\omega^{3}),
\label{eigenvalue}
\end{gather}
where $h(l)$ is given by
\begin{gather}
h(l)=\frac{2(l^{2}-m^{2})(l^{2}-s^{2})^{2}}{(2l-1)l^{3}(2l+1)}.
\end{gather}
For the nonhomogeneous case, we expand $\psi_{(s)}$ and $T_{(s)}$ in terms of ${}^{}_{s}S^{a\omega}_{lm}(\theta,\varphi)$ as
\begin{align}
\psi_{(s)}&=\int d\omega \sum_{l,m}R^{(s)}_{lm\omega}(r){}^{}_{s}S^{a\omega}_{lm}(\theta,\varphi)e^{-i\omega t},
\\
-2\Sigma T_{(s)}&=\int d\omega \sum_{l,m}G^{(s)}_{lm\omega}(r){}^{}_{s}S^{a\omega}_{lm}(\theta,\varphi)e^{-i\omega t}.
\end{align}
Then $R^{(s)}_{lm\omega}(r)$ satisfies
\begin{align}
&L^{-s}\frac{d}{dr}\left(L^{s+1}\frac{dR^{(s)}_{lm\omega}}{dr}\right)
\notag\\
&\qquad {}
+\left[\frac{K^{2}-isKL'}{L}+\frac{1}{2}(|s|+s)(L''-2)+4is\omega r-\tilde{\boldsymbol{\lambda}}_{(s)}\right]R^{(s)}_{lm\omega}
=G^{(s)}_{lm\omega}.
\label{radial2}
\end{align}
where $\tilde{\boldsymbol{\lambda}}_{(s)}$ is defined by
\begin{gather}
\tilde{\boldsymbol{\lambda}}_{(s)}=\boldsymbol{\lambda}_{(s)}-2ma\omega+a^{2}\omega^{2}.
\end{gather}
We again note that in the case of $L(r)=r^{2}-2Mr+a^{2}$, \eqref{radial2} reduces to the Teukolsky radial equation
with the spin $s$ on the background of the Kerr spacetime.
We can also perform a similar transformation as \eqref{redef2} as
\begin{gather}
\hat{R}^{(s)}_{lm\omega}=L^{\frac{1}{2}(|s|+s)}R^{(s)}_{lm\omega}, \quad
\hat{G}^{(s)}_{lm\omega}=L^{\frac{1}{2}(|s|+s)}G^{(s)}_{lm\omega},
\end{gather}
and then $\hat{R}^{(s)}_{lm\omega}$ satisfies
\begin{gather}
L^{|s|}\frac{d}{dr}\left(L^{-|s|+1}\frac{d\hat{R}^{(s)}_{lm\omega}}{dr}\right)
+\left(\frac{K^{2}-isKL'}{L}+4is\omega r-\hat{\boldsymbol{\lambda}}_{(s)}\right)\hat{R}^{(s)}_{lm\omega}
=\hat{G}^{(s)}_{lm\omega}.
\label{radial3}
\end{gather}

\section{Singularity structure of radial equation}

The structure of the wave equation obtained in the previous section strongly depends on the behavior of $L(r)$.
In particular the large-$r$ and the small-$r$ behaviors are important. We introduce the dimensionless variable $z=\omega r$,
and define
\begin{gather}
\tilde{L}(z)=\omega^{2}L(z/\omega).
\end{gather}
It is in general difficult to obtain the exact form of $\tilde{L}(z)$ except some special cases, such as the vacuum.
What one can usually perform is to find the large-$z$ expansion of $\tilde{L}(z)$ using the post-Newtonian or the post-Minkowskian method.
For the leading term, $\tilde{L}(z)\sim z^{2}$ by assuming the asymptotic flatness, and then $\tilde{L}(z)$ may be expanded as
\begin{gather}
\tilde{L}(z)=z^{2}+a_{-1}z+a_{0}+\frac{a_{1}}{z}+\cdots+\frac{a_{n}}{z^{n}}+\cdots.\,
\label{expansion}
\end{gather}
where $a_{j}$ $(j=-1, 0, 1, \cdots)$ are constant.
Since the above form is hard to handle,
we here terminate the expansion up to the term $a_{n}z^{-n}$, and express $\tilde{L}(z)$ as
\begin{gather}
\tilde{L}(z)\sim z^{-n}P_{n+2}(z),
\label{ansatz}
\end{gather}
where $P_{n+2}(z)$ denotes a monic polynomial of $z$ with the degree $n+2$.
One can assume $a_{n}\neq 0$ without loss of generality, because otherwise one can shift $n$ to $n-1$ or less.
By this procedure we will always take the minimum $n$.
In the case of $n=0$, the background reduces to the Kerr spacetime if the signs of the coefficients $a_{-1}$ and $a_{0}$ are appropriate.
The Schwarzschild spacetime corresponds to the case of $n=-1$. The flat spacetime corresponds to $n=-2$.

We also parametrize $P_{n+2}(z)$ by its roots $\alpha_{k}$ $(k=1,2,\ldots,n+2)$ as%
\footnote{There is always a possibility that we have some artificial zeroes
from the termination of the expansion, which may tend to disappear or to move into the infinity,
when we include the higher order of the expansion \eqref{expansion}. }
\begin{gather}
P_{n+2}(z)=\prod_{k=1}^{n+2}(z-\alpha_{k}),
\label{roots}
\end{gather}
where $\alpha_{k}$ are in general complex.
One can find that none of $\alpha_{k}$ vanishes from $a_{n}\neq 0$.
Substituting \eqref{ansatz} into the transformed homogeneous radial equation \eqref{radial0}, we have
\begin{align}
&\frac{d^{2}\hat{R}}{dz^{2}}+(-|s|+1)\left(\frac{P'_{n+2}}{P_{n+2}}-\frac{n}{z}\right)\frac{d\hat{R}}{dz}
\notag\\
&\qquad
+\left[\frac{z^{2n}}{(P_{n+2})^{2}}\hat{K}^{2}-is\hat{K}\frac{z^{n}}{P_{n+2}}\left(\frac{P'_{n+2}}{P_{n+2}}-\frac{n}{z}\right)
+\frac{z^{n}}{P_{n+2}}(4isz-\hat{\boldsymbol{\lambda}}_{(s)})
\right]\hat{R}=0,
\label{diffeq1}
\end{align}
where $\hat{K}$ is defined by
\begin{gather}
\hat{K}=\omega K=z^{2}-ma\omega+a^{2}\omega^{2}.
\end{gather}
If all of  $\alpha_{k}$ are the single roots,
then from \eqref{diffeq1} one can find that the coefficient of $d\hat{R}/dz$ has at most the single poles at $z=0$, $\alpha_{k}$,
and that of $\hat{R}$ has at most the double poles at the same points, which implies that $z=0$, $\alpha_{k}$ are the regular singular points.
If some of $\alpha_{k}$ are the multiple roots, then those become the irregular singular points.
Hereafter we assume that all of  $\alpha_{k}$ are the single roots for simplicity, unless it is specified.
On the other hand, $z=0$ always appears as the regular singularity, regardless of the value of $n$.
One can also examine the singularity structure of the equation around the infinity $z=\infty$,
which can be studied by applying the M\"{o}bius transformation $z=1/w$, and then by examining the behaviour around $w=0$.
One can find that $z=\infty$ is in turn the irregular singular point with the Poincar\'{e} rank \cite{Heun} unity.

The structure of the singularities in the above can for example be compared with that of the radial equation on the background of
the Kerr spacetime, which is the original Teukolsky equation.
In the (non-extremal) Kerr spacetime, $\tilde{L}(z)$ takes the form of
\begin{align}
\tilde{L}(z)&=z^{2}-2M\omega z +(a\omega)^{2}=(z-\alpha_{+})(z-\alpha_{-}),
\label{Kerr}
\\
\alpha_{\pm}&=M\omega \pm \sqrt{(M\omega)^{2}-(a\omega)^{2}},
\end{align}
where the roots $\alpha_{+}$ and $\alpha_{-}$ correspond to the outer and the inner horizon for $a<M$, respectively.
The origin $z=0$ does not appear as the singularity because of $n=0$, and hence the radial equation \eqref{diffeq1} has two regular singularities
at $z=\alpha_{\pm}$ and one irregular singularity at $z=\infty$ with the Poincar\'{e} rank unity.
Then the equation is classified as an example of the confluent Heun equation \cite{Heun}.
In the extremal case $a=M$, the roots $\alpha_{+}$ and $\alpha_{-}$ of $\tilde{L}(z)$ coincide and become the double root,
which gives the irregular singularity. Because of that, the radial equation is classified as an example of the doubly confluent Heun equation \cite{Heun}.
We note that instead of the parametrization \eqref{ansatz} which is the termination of the expansion \eqref{expansion},
one can also use the Pad\'{e} approximation as
\begin{gather}
\tilde{L}(z)\sim\frac{P_{n+2}(z)}{Q_{n}(z)},
\label{Pade}
\end{gather}
where $Q_{n}(z)$ is a monic polynomial with the degree $n$. \eqref{ansatz} is a special case of the above as $Q_{n}(z)=z^{n}$.
By substituting \eqref{Pade} into the radial equation, one can find that the roots%
 $\beta_{k'}$ of $Q_{n}(z)$ appear
as at most the single poles in the coefficient of $d\hat{R}/dz$, and as at most the double poles in that of $\hat{R}$.
Then $z=\beta_{k'}$ always give the regular singularities of the equation, regardless of their multiplicities, where $k'$ runs from 1 to the number of the distinct roots.

Next we will examine the behaviour of the solution to \eqref{diffeq1} around each singularity using the Frobenius-like method,
following Teukolsky \cite{Teukolsky:1973ha}.
We will first consider \eqref{ansatz} instead of the Pad\'{e} approximation \eqref{Pade} for simplicity.
We introduce the tortoise coordinate $z^{\ast}$ as
\begin{gather}
\frac{dz^{\ast}}{dz}=\frac{z^{2}+a^{2}\omega^{2}}{\tilde{L}(z)},
\label{tortoise1}
\end{gather}
where one can apply the partial fraction decomposition to the right hand side as
\begin{gather}
\frac{z^{2}+a^{2}\omega^{2}}{\tilde{L}(z)}=
\frac{z^{n}(z^{2}+a^{2}\omega^{2})}{\prod_{k}(z-\alpha_{k})}
=1+\sum_{k}\frac{\alpha_{k}^{n}(\alpha_{k}^{2}+a^{2}\omega^{2})}{P'_{n+2}(\alpha_{k})}\frac{1}{z-\alpha_{k}}.
\end{gather}
Then \eqref{tortoise1} can be integrated as
\begin{gather}
z^{\ast}=z+\sum_{k}\frac{\alpha_{k}^{n}(\alpha_{k}^{2}+a^{2}\omega^{2})}{P'_{n+2}(\alpha_{k})}\ln(z-\alpha_{k})+\textrm{const}.
\label{tortoise2}
\end{gather}
We also introduce the function $Y(z)$ as
\begin{gather}
Y=\tilde{L}^{-|s|/2}(z^{2}+a^{2}\omega^{2})^{1/2}\hat{R}=\tilde{L}^{s/2}(z^{2}+a^{2}\omega^{2})^{1/2}R.
\end{gather}
$Y(z)$ satisfies the following differential equation:
\begin{gather}
\frac{d^{2}Y}{dz^{\ast 2}}+\left[\frac{\hat{K}^{2}}{(z^{2}+a^{2}\omega^{2})^{2}}-\frac{is\hat{K}\tilde{L}'}{(z^{2}+a^{2}\omega^{2})^{2}}
+\frac{\tilde{L}(4isz-\hat{\boldsymbol{\lambda}}_{(s)})}{(z^{2}+a^{2}\omega^{2})^{2}}-G^{2}-\frac{dG}{dz^{\ast}}\right]Y=0,
\label{diffeq2}
\end{gather}
where $G(z)$ is defined by
\begin{gather}
G(z)=-\frac{|s|\tilde{L}'(z)}{2(z^{2}+a^{2}\omega^{2})}+\frac{z\tilde{L}(z)}{(z^{2}+a^{2}\omega^{2})^{2}}.
\end{gather}
When $z$ is large, \eqref{diffeq2} is approximated as
\begin{gather}
\frac{d^{2}Y}{dz^{\ast 2}}+\left(1+\frac{2is}{z}\right)Y \sim 0,
\end{gather}
and then the behaviour of the solution near the infinity $z=\infty$ is
$Y \sim z^{\mp s}e^{\pm iz^{\ast}}$, namely,
\begin{gather}
R(z) \sim C_{\textrm{in}}z^{-1}e^{-iz^{\ast}} + C_{\textrm{out}}z^{-(2s+1)}e^{iz^{\ast}},
\label{sol1}
\end{gather}
where $C_{\textrm{in}}$ and $C_{\textrm{out}}$ are constant.
We note that the behaviour of \eqref{sol1} is the same as that for the Teukolsky equation due to the asymptotic flatness.

On the other hand, around one of the root $\alpha_{k}$ of $\tilde{L}(z)$, \eqref{diffeq1} is approximated as
\begin{gather}
\frac{d^{2}Y}{dz^{\ast 2}}+\left[p-\frac{is\tilde{L}'(\alpha_{k})}{2(\alpha_{k}^{2}+a^{2}\omega^{2})}\right]^{2}Y \sim 0,
\end{gather}
where $p$ is defined by
\begin{gather}
p=1-\frac{ma\omega}{\alpha_{k}^{2}+a^{2}\omega^{2}}.
\end{gather}
Then the solution to \eqref{diffeq2} near $z=\alpha_{k}$ behaves as
\begin{gather}
Y(z) \sim \exp\left[\pm i\left(p-\frac{is\tilde{L}'(\alpha_{k})}{2(\alpha_{k}^{2}+a^{2}\omega^{2})}\right)z^{\ast}\right].
\end{gather}
By the use of \eqref{tortoise2}, the above can be rewritten as
\begin{align}
Y(z) &\sim e^{\pm ipz^{\ast}}\exp\left[\pm \frac{s}{2} \frac{\tilde{L}'(\alpha_{k})}{\alpha_{k}^{2}+a^{2}\omega^{2}}
\frac{\alpha_{k}^{n}(\alpha_{k}^{2}+a^{2}\omega^{2})}{P'_{n+2}(\alpha_{k})}\ln(z-\alpha_{k})\right]
\notag\\
&\sim (z-\alpha_{k})^{\pm s/2} e^{\pm ipz^{\ast}}
\notag\\
&\sim \tilde{L}^{\pm s/2} e^{\pm ipz^{\ast}}.
\end{align}
Namely,
\begin{gather}
R(z) \sim B_{\textrm{in}}e^{-ipz^{\ast}} + B_{\textrm{out}}\tilde{L}^{-s}e^{ipz^{\ast}}
\label{sol2}
\end{gather}
where $B_{\textrm{in}}$ and $B_{\textrm{out}}$ are constant.
We note that \eqref{sol2} directly reduces to the Kerr case, when \eqref{Kerr} is satisfied. In particular, the exponents
(zero for $e^{-ipz^{\ast}}$ and $-s$ for $e^{ipz^{\ast}}$) are the same as the those in the Kerr case.

Finally, the behaviour around $z=0$ is nontrivial only when $n$ is positive.
In the case of the negative $n$, it reduces to \eqref{sol2} with $\alpha_{k}\to 0$.
The structure of \eqref{diffeq1} near $z = 0$ is simpler than that of \eqref{diffeq2} as
\begin{gather}
\frac{d^{2}\hat{R}}{dz^{2}}-\frac{n(-|s|+1)}{z}\frac{d\hat{R}}{dz}+C\hat{R} \sim 0,
\end{gather}
where the constant $C$ is given by
\begin{gather}
C=
\begin{cases}
-is(a^{2}\omega^{2}-ma\omega)a_{1}^{-1} &  \text{for $n=1$}
\\
0 & \text{for $n\ge 2$}.
\end{cases}
\end{gather}
By assuming the power solution $\hat{R}\sim z^{\xi}$, we have $\xi=0$, $n(1-|s|)+1$. The special case yields when $n(1-|s|)+1=0$,
namely $n=1$, $|s|=2$. In this case we have the logarithm $\hat{R}\sim \ln z$ as well as the constant solution.
Then the solution $R(z)$ near $z=0$ behaves as
\begin{gather}
R(z) \sim
\begin{cases}
A_{1}z^{1+s/2}\ln z + A_{2}z^{1+s/2} & \text{for $n=1$, $|s|=2$}
\\
\tilde{A}_{1}z^{n+1+n(-|s|+s)/2}+\tilde{A}_{2}z^{n(|s|+s)/2} & \text{otherwise},
\end{cases}
\label{sol3}
\end{gather}
where $A_{1}$, $A_{2}$, $\tilde{A}_{1}$ and $\tilde{A}_{2}$ are constant.
We note that \eqref{sol3} can also be used for the case of the Pad\'{e} approximation \eqref{Pade} by the following replacement:
\begin{gather}
z \to z-\beta_{k'}, \quad n \to n_{k'},
\end{gather}
where $\beta_{k'}$ is one of the root of $Q_{n}(z)$ in \eqref{Pade} and $n_{k'}$ is the multiplicity of $\beta_{k'}$.
On the other hand, \eqref{sol1} and \eqref{sol2} hold for the case of the Pad\'{e} approximation as well.

\section{Summary and discussion}

In this paper we have studied the wave equations with the various spins on the background of the deformed Kerr metric \eqref{metric1},
By introducing the spin variable $s$, we have unified those equations using $s$ itself and $|s|$ as \eqref{unified2}.
We have found that taking the separable gauge \eqref{gauge03} and \eqref{gauge04}, the unified wave equation \eqref{unified2}
becomes simpler than that on the background of the general spherically symmetric spacetime \cite{Guo:2023hdn}. After the substitution of the
Newman-Penrose quantities, the explicit partial differential equation admits the separation of the variables.
The angular part of the separated equation \eqref{angular} gives the spin-weighted spheroidal harmonics, as expected, and
the radial part \eqref{radial} gives the natural extension of the Teukolsky equation.

We have also examined that the existence of the singularity in the radial equation and the behaviour of the solution around each singularity,
by the use of the large-$z$ expansion \eqref{expansion} or Pad\'{e} approximation \eqref{Pade}. We have found that the singularity exists
at the infinity, the zeroes and the poles of $\tilde{L}(z)$. The behaviours of the solution around the infinity
is the same as that of the Teukolsky equation, and behaviour of the solution around the zeroes of $\tilde{L}(z)$ is very similar.

Here our study is restricted to the case that the conformal factor $A(r,\theta)$ in \eqref{metric2} is identically unity,
otherwise the components of the Ricci tensor become nonvanishing and the system becomes complicated except for some special case,
such as the spherically symmetric background. However whether some components vanish or do not vanish may depend on the choice of the tetrad basis.
We have here assumed the specific form of the basis \eqref{tetrads1}, which is the natural extension of the standard choice for the Kerr case.
It would be useful to find the better expression for the tetrad basis for the case of $A(r,\theta)\neq 1$.
For this problem, the Pleba\'{n}ski-Segre classification \cite{Plebanski, Segre} (see also \cite{Stephani:2003tm}),
the analogue of the Petrov classification for the (trace-free) Ricci tensor, would be important.
The Pleba\'{n}ski-Segre classification may also be helpful to find the non-vacuum extension of the Kinnersley metric \cite{Kinnersley:1969zza}
which gives the complete classification of the type D vacuum metrics.

We have also found that the behaviour of the solution to the homogeneous radial equation at each singularity does not depend on the detail of
$\tilde{L}(z)$, under the assumption that all of the roots of $\tilde{L}(z)$ are the single roots,
which is reasonable because we are now considering the local behaviours. The detail of $\tilde{L}(z)$ is then important for the
global behaviour of the solution, in particular the relation between the coefficients among \eqref{sol1}, \eqref{sol2} and \eqref{sol3},
called the connection problem,
which is also important to solve the scattering problem and to obtain the quasi-normal frequency
\cite{Arbey:2021yke, Cardoso:2021wlq, Cardoso:2022whc, Destounis:2022obl, Figueiredo:2023gas}.
Recently, there are huge developments \cite{Bonelli:2021uvf, Bonelli:2022ten, Fucito:2023afe, Bautista:2023sdf}
to calculate the scattering amplitude by solving the connection problem by the use of the
relation between the two-dimensional (Liouville or Toda) conformal field theories (CFTs) and the $\mathcal{N}=2$ supersymmetric gauge theories
\cite{Seiberg:1994rs, Seiberg:1994aj, Nekrasov:2002qd, Nekrasov:2003rj, NS},
called the Alday-Gaiotto-Tachikawa (AGT) conjecture \cite{Alday:2009aq, Gaiotto:2009ma}.
A similar analysis may be able to be performed in the current case of the non-vacuum background.
It is important the connection coefficients for the solution to the differential equation with more singularities, which correspond to
more insertions of the primary operators in the CFTs and more matter hypermultiplets in the $\mathcal{N}=2$ supersymmetric gauge theories.

\section*{Acknowledgements}
This work was supported in part by the National Natural Science Foundation of China (Grant No. 11973025).



\end{document}